\documentstyle[aps,preprint]{revtex}

\newcommand{\be}{\begin{equation}}
\newcommand{\ee}{\end{equation}}
\newcommand{\ba}{\begin{eqnarray}}
\newcommand{\ea}{\end{eqnarray}}

\begin{document}
%\twocolumn[\hsize\textwidth\columnwidth\hsize\csname@twocolumnfalse\endcsname

\draft
\date{\today}
\title{Generalized BFT Formalism of Electroweak Theory in the Unitary Gauge}
\author{Yong-Wan Kim\footnote{e-mail:~ywkim@physics3.sogang.ac.kr},
Young-Jai Park\footnote{e-mail:~yjpark@ccs.sogang.ac.kr}
and Sean J. Yoon\footnote{
e-mail:~yoonsj@mail.lgcit.com,\\ 
Permanent address:~LG Corporate Institute of Technology, Seoul
137-724, Korea}}
\address{Department of Physics, Sogang University, 
C.P.O.Box 1142,
Seoul 100-611, Korea}
\maketitle
\begin{abstract}
We systematically embed the SU(2)$\times$U(1) 
Higgs model in the unitary gauge into a fully gauge invariant theory
by following the generalized BFT formalism.
We also suggest a novel path how to get 
a first-class Lagrangian directly from the original 
second-class one using the BFT fields.
\end{abstract}


%\narrowtext
%\vskip1pc]

A phenomenological example of constrained systems \cite{Di,HT}
is provided by the SU(2)$\times$U(1) Higgs model with 
spontaneous symmetry breakdown whose quantization
is usually carried out in the so called ``unitary'' gauge. 
The model in this gauge is characterized by
both of second- and first-class constraints in the language of Hamiltonian
formalism. 
Since the strong implementation of second-class constraints 
generally leads to non-polynomial field dependent
Dirac brackets which may pose problems, 
one can circumvent them associated with this
non-polynomial dependence by embedding second-class constrained system
into first-class one in an extended phase space through
the formalism of Batalin {\it et al} (BFT) \cite{BFT}.
Now, while preserving the Poisson structure, 
we can successfully quantize the system 
implementing first-class constraints on physical states. 
Recently, we have introduced BFT fields \cite{KK} 
through which the construction
of physical observables such as Hamiltonian is much easier than the
direct adoption of BFT method. When used BFT fields in 
second-class constrained system, there is an elegant one-to-one
correspondence to first-class constrained system in an extended phase
space. Moreover, the use of BFT fields makes it possible to analyze a
rather complicated non-abelian constrained system \cite{BB,ON} conveniently.
On the other hand, we have studied the spontaneously broken
abelian U(1) \cite{KP} and non-abelian SU(2) \cite{KPR} 
Higgs models, which are of fully 
second-class constraints due to the completely broken symmetry, as toy models.

In this paper we analyze the non-abelian SU(2)$\times$U(1) 
Higgs model in the unitary gauge by following the generalized BFT 
procedure \cite{BFT,KK}. This real phenomenonlogical model is 
highly non-trivial contrast to 
the U(1) and SU(2) models because there still remains U(1)$_{em}$ symmetry 
after spontaneous symmetry breaking.
It also needs to embed first-class constraints as well as second-class
constraints in this SU(2)$\times$U(1) Higgs model in order to keep the 
consistency and simplicity of constraint algebra
as shown it later.

Starting from the second-class Lagrangian, 
we construct an effectively first-class constrained system.
We then show that the results by using BFT fields coincide with those 
obtained by gauging the second-class Lagrangian and performing a suitable 
canonical transformation. 
We also suggest an economic novel path at the classical level 
to obtain the first-class Lagrangian from the second-class one.

%%%%%%%%%%%%%%%%%%%%%%%%%%%%%%%%%%%%%%%%%%%%%%%%%%%%%%%%%%%%%%%%%%%%%%%%%%

Let us consider the non-abelian SU(2)$\times$U(1) Higgs model 
in the unitary gauge, 
which describes the bosonic part of the Weinberg-Salam model \cite{WeSa}, 
\ba\label{1}
{\cal L}_u = -{1 \over 4} F_{\mu\nu}^a F^{\mu\nu a}
     - {1 \over 4} G_{\mu\nu}G^{\mu\nu} 
     + {1\over 8} {\rho}^2 ( g'^2 B_\mu B^\mu
        -2 g g' B_\mu A^{\mu 3} + g^2 A_\mu^a  A^{\mu a})
     +{1 \over 2}\partial_\mu \rho \partial^\mu \rho + V(\rho),
\ea
where $V(\rho)$ is the Higgs potential, 
$V(\rho) = {\mu^2 \over 2} (\rho+ v)^2 - {\lambda \over 4} (\rho+ v)^4$,
with the vacuum expectation value $v$,
and the $g'$ and $g$ denote the U(1) and SU(2) coupling constants, 
respectively. 
The field strength tensors are
$F_{\mu\nu}^a =\partial_\mu A_{\nu}^a -\partial_\nu A_{\mu}^a
            + g \epsilon^{abc} A_{\mu}^b A_{\nu}^c, ~(a=1,2,3)$ and
$G_{\mu\nu}=\partial_\mu B_{\nu}-\partial_\nu B_{\mu}$.
The momenta canonically conjugate to $A^{0a}$, $A^{ia}$, 
$B^0$, $B^i$, and $\rho$ in the Hamiltonian formalism are 
given by $\pi^a_0=0$, $\pi^a_i=F^a_{i0}$, $p_0=0$, $p_i=G_{i0}$,
and $\pi_\rho=\dot \rho$, respectively. 
We thus have the primary constraints $\pi^a_0\approx0$ and $p_0\approx0$.
The canonical Hamiltonian density associated with the Lagrangian 
(\ref{1}) is found to be
\ba\label{2}
{\cal H}_C &=& {1\over 2} (\pi^a_i)^2 
       + {1\over 2} (p_i)^2 
       + {1\over 2} (\pi_\rho)^2
       + {1 \over 4} F^a_{ij} F^{ija}
       + {1 \over 4} G_{ij} G^{ij}
       + {1\over 8} {\rho}^2  ( g'^2 (B^i)^2 
             - 2 g g' B^i  A^{i3} + g^2 (A^{ia})^2 )
  \nonumber\\      
  &&   + {1\over 2} (\partial_i \rho)^2
             - V(\rho)
       - {1\over 8} {\rho}^2 ( g'^2 (B^0)^2 
          -2 g g' B^0  A^{03} + g^2 (A^{0a})^2 )
          - A^{0a} ({\cal D}^i \pi_i)^a - B^0 \partial^i p_i ,
\ea
where $({\cal D}^i \pi_i)^a 
= \partial^i \pi^a_i +g  \epsilon^{abc} A^{ib} \pi^c_i$.
Since persistency in time of the primary constraints 
leads to further constraints, 
this system is described by the set of eight constraints as
\ba\label{3}
&& \omega_1 =  p_0, \nonumber \\
&& \omega_2=\partial^i p_i  - {1\over 4} {\rho}^2 g g' A^{03}
            + {1\over 4} {\rho}^2 g^{'2} B^0,\nonumber \\
&&\omega^a_3=\pi^a_0,\nonumber \\
&& \omega^a_4=({\cal D}^i \pi_i)^a + {1\over 4} {\rho}^2 g^2 A^{0a}
            - {1\over 4} {\rho}^2 g g' B^0 \delta^{a3}.  
\ea
The corresponding constraints algebra is given by
\ba\label{4}
\Sigma_{ij}&=&\{\omega_i(x), \omega_j(y)\}
                                            \nonumber \\ 
       &=& \left(\begin{array}{cc}
                \left(\begin{array}{cc} 
                      0 & -{1\over 4} {\rho}^2 g^{'2} \\
                      {1\over 4} {\rho}^2 g^{'2} &   0 
                      \end{array}\right) _{U(1)} &
                      \left(\begin{array}{cc} 
                         0 & {1\over 4} {\rho}^2 g g' \delta^{a3} \\
                         - {1\over 4} {\rho}^2 g g' \delta^{a3}  &   0 
                      \end{array}\right) _{mixed} \\                         
                      \left(\begin{array}{cc} 
                         0 & {1\over 4} {\rho}^2 g g' \delta^{a3} \\
                         - {1\over 4} {\rho}^2 g g' \delta^{a3}  &   0 
                      \end{array}\right) _{mixed} &
                      \left(\begin{array}{cc} 
                         0 & -{1\over 4} {\rho}^2 g^2  \delta^{ab}  \\
                         {1\over 4} {\rho}^2 g^2  \delta^{ab}   &  
                         g \epsilon^{abc}\left({\cal D}^k \pi_k \right)^c
                      \end{array}\right) _{SU(2)}  
                   \end{array}\right) \delta^3(x-y).\ea
Each block of the matrix  represents the U(1) \cite{KP} 
and SU(2) Higgs models \cite{KPR}
with additional mixed components in the off-diagonal part.   
But, the matrix (\ref{4}) is of zero determinant, which means that 
among the constraints (\ref{3}) there still exist first-class ones
related to the unbroken symmetry. 
In order to efficiently extract out first-class constraints,
let us redefine the constraints (\ref{3}) as 
\ba\label{5}
\Omega^a_1 &\equiv& \omega^a_3 ~=~ \pi^a_0, \nonumber \\
\Omega^a_2 &\equiv& \omega^a_4 ~=~
                ({\cal D}^i \pi_i)^a + {1\over 4} {\rho}^2 g^2 A^{0a}
               - {1\over 4} {\rho}^2 g g' B^0 \delta^{a3}, \\
\label{6}
T_1&\equiv& g \omega_1 + g' \omega^3_3 
        ~=~g p_0 + g' \pi^3_0, \nonumber \\ 
T_2&\equiv&  g \omega_2 +  g' \omega^3_4 + 
              {4\over {\rho^2 g^2}} [g g' ({\cal D}^i \pi_i)^2  \omega^1_3  
                                   - g g' ({\cal D}^i \pi_i)^1  \omega^2_3 ]
              \nonumber \\
&=&  g \partial^i p_i +  g' (D^i \pi_i)^3 + 
               {4\over {\rho^2 g^2}} [g g' ({\cal D}^i \pi_i)^2  \pi^1_0
                 - g g' ({\cal D}^i \pi_i)^1  \pi^2_0] .
\ea
Then, the constraints  $\Omega^a_1$ and $\Omega^a_2$
become second-class as 
\be\label{7}
\{\Omega^a_i(x),\Omega^b_j(y)\}=
                  \Delta^{ab}_{ij}(x,y)=\left(\begin{array}{cc}
                           0 & -{1\over 4} {\rho}^2 g^2  \delta^{ab}  \\
                           {1\over 4} {\rho}^2 g^2  \delta^{ab}   &  
                           g \epsilon^{abc}\left({\cal D}^k \pi_k \right)^c
                      \end{array}\right)\delta^3(x-y),\ee
while the $T_1$ and $T_2$ are first-class as
\ba\label{8}
\{T_2 (x),\Omega^1_2 (y)\}&=& 
      -{4\over {\rho^2}} g' ({\cal D}^i \pi_i)^3  \Omega^1_1 \delta^3(x-y)
      \approx 0,   \nonumber \\
\{T_2 (x),\Omega^2_2 (y)\}&=& 
      -{4\over {\rho^2}} g' ({\cal D}^i \pi_i)^3  \Omega^2_1 \delta^3(x-y)
      \approx 0,  \nonumber \\
\{T_2 (x),\Omega^3_2 (y)\}&=& 
      {4\over {\rho^2}} g' [({\cal D}^i \pi_i)^1  \Omega^1_1+
             ({\cal D}^i \pi_i)^2  \Omega^2_1] \delta^3(x-y)
             \approx 0, \nonumber \\
\{T_1 (x),\Omega^a_i (y)\} &=& \{T_i (x), T_j (y)\} = 0.
\ea 
The SU(2)$\times$U(1) Higgs model in the unitary gauge
thus has two first- as well as six second-class constraints. 
As is well known, after breaking the SU(2)$\times$U(1) symmetry 
spontaneously, the SU(2)$\times$U(1) symmetry is broken
into the combined symmetry U(1)$_{em}$.
As a result, the vector fields, 
$W^{\pm}_{\mu} \left(=(A^1_\mu \mp i A^2_\mu)/\sqrt{2}\right)$
and $Z_\mu (={\rm cos}\theta A^3_\mu - {\rm sin} \theta B_\mu, 
~{\rm tan}\theta= {g'/ g})$ have acquired masses equal to 
$M^2_W (\rho) = {1 \over 4} \rho^2 g^2$ and
$M^2_Z (\rho) = {1 \over 4} \rho^2 (g^2 + g'^2)$,
while $A_\mu(={\rm sin}\theta A^3_\mu 
+ {\rm cos} \theta B_\mu)$ has remained massless.
In fact, the first-class constraints (\ref{6}) describe this
residual U(1)$_{em}$ symmetry, {\it i.e.}, 
the existence of the massless gauge fields.  

We now convert this system of the second- $\Omega^a_i$
and first-class constraints $T_i$
into a completely equivalent first-class system
at the expense of additional degrees of freedom. 
It is important to note that we should embed the first-class constraints
as well as the second-class constraints. Embedding only the second-class
constraints $\Omega^a_i$ in general does not preserve the first-class
algebra as it is clear in (\ref{8}).

In order to embed the second-class constraints (\ref{5}) by 
following the BFT method \cite{BFT}, 
we introduce auxiliary fields $\Phi^{1a}$ and $\Phi^{2a}$ corresponding to 
$\Omega^a_1$ and  $\Omega^a_2$ of symplectic structure
\be\label{9}
\left\{\Phi^{ia}(x),\Phi^{jb}(y)\right\}=\omega^{ij}_{ab}(x,y)
         =\epsilon^{ij}\delta_{ab}\delta^3(x-y),~~(i,j=1,2).
\ee
The effective first-class constraints $\tilde\Omega^a_i$ 
are now constructed as a power series in the auxiliary fields,
$\tilde\Omega^a_i=\Omega^a_i+\sum^\infty_{n=1}\Omega^{(n)a}_i$,
where $\Omega^{(n)a}_i(n=1,...,\infty)$ are homogeneous polynomials in the  
auxiliary fields $\{\Phi^{jb}\}$ of degree $n$. These 
will be determined by the  
requirement that the constraints $\tilde\Omega^a_i$ be strongly involutive:
\be\label{10}
\left\{\tilde\Omega^a_i(x),\tilde\Omega^b_j(y)\right\}=0.\ee
Making a general ansatz
\be\label{11}
\Omega^{(1)a}_i(x)=\int d^3y X^{ab}_{ij}(x,y)\Phi^{jb}(y), \ee
and substituting (\ref{11}) into (\ref{10}) leads to the condition of
\be\label{12}
\int d^3 z X^{ac}_{ik}(x,z)\epsilon^{k\ell}X^{bc}_{j\ell}(y,z)
=-\Delta^{ab}_{ij}(x,y).\ee
Then, Eq. (\ref{12}) has a solution of
\be\label{13}
X^{ab}_{ij}(x,y)=\left(\begin{array}{cc}
{1\over 4} {\rho}^2 g^2 \delta^{ab}&0\\
-{1\over 2}g \epsilon^{abc}({\cal D}^k\pi_k)^c&\delta^{ab}\end{array}\right)
\delta^3(x-y).\ee

>From the symplectic structure of (\ref{9}), 
we may identify the auxiliary fields with canonically
conjugated pairs. We make this explicit by adopting the notation,
$(\Phi^{1a}, \Phi^{2a}) \Rightarrow (\theta^a, \pi^a_\theta)$. 
Substituting (\ref{13}) into (\ref{11}) and  
iterating this procedure one finds the strongly involutive first-class  
constraints to be given by
\ba\label{14}
&&\tilde\Omega^a_1=\pi^a_0+{1\over 4} {\rho}^2 g^2  \theta^a,\nonumber \\
&&\tilde\Omega^a_2= V^{ab}(\theta)({\cal D}^i\pi_i)^b +
{1\over 4} {\rho}^2 g^2 A^{0a} - {1\over 4} {\rho}^2 g g' B^0 \delta^{a3}
+ \pi_\theta^a.
\ea
Here,
$V(\theta)=\sum^\infty_{n=0}{{(-1)^n}\over{(n+1)!}}
({\it ad}\, \theta)^n$,
and
$\left({\it ad}\, \theta \right)^{ab}= g\epsilon^{acb}\theta^c$,
where ${\it ad} \,\theta=g \theta^aT^a$ with $T^c_{ab}=\epsilon^{acb}$
denotes the Lie algebra-valued field 
$\theta$ in the adjoint representation.
This completes the construction of the effective first-class constraints
corresponding to the second-class ones.

%%%%%%%%%%%%%%%%%%%%%%%%%%%%%%%%%%%%%%%%%%%%%%%%%%%%%%%%%%%%%%%%%%%%%%%%%%%

The construction of effective 
first-class Hamiltonian $\tilde H$ can be also obtained
along similar lines as in the case of the constraints, 
{\it i.e.,} by representing it as  
a power series in the auxiliary fields and requiring
$\{\tilde\Omega^a_i,\tilde H\}=0$ subject to the condition 
$\tilde H[{\cal J};\theta^a = \pi^a_\theta = 0]=H_C$,
where ${\cal J}$ denotes collectively the variables 
$(A^{\mu a},\pi^a_\mu,B^{\mu},p_\mu,\rho,\pi_\rho)$ 
of the original phase space.
However, we shall follow a  novel
path \cite{KK,BB,KR} by noting that any 
functional of first-class fields will also be first-class.
We require BFT fields $\tilde {\cal J}$ corresponding to ${\cal J}$
in the extended phase space to be 
strongly involutive with the effective
first-class constraints $\tilde \Omega^a_i$, {\it i.e.},
$\{\tilde\Omega^a_i,\tilde{\cal J} \}=0$,
which leads us to the identification $\tilde H=H_C[\tilde {\cal J}]$.
The BFT fields $\tilde{\cal J}$ are now obtained 
as a power series in the  
auxiliary fields $(\theta^a, \pi^a_\theta)$, 
whose iterative solutions lead to the following compact infinite series as
\ba\label{15}
\tilde A^{0a}&=& A^{0a}+\frac{4}{g^2 \rho^2}\pi_\theta^a-
\frac{4}{g^2 \rho^2}\left(U^{ab}(\theta)-V^{ab}(\theta)\right)
({\cal D}^i\pi_i)^b,\nonumber\\
\tilde A^{ia}&=& U^{ab}(\theta) A^{ib}+V^{ab}(\theta)\partial^i\theta^b,
\nonumber \\
\tilde\pi^a_\mu &=& \left( \pi^a_0+\frac{1}{4} g^2 \rho^2\theta^a,~
             U^{ab}(\theta)\pi_i^b \right), \nonumber\\
\tilde B^\mu &=& B^\mu,~~~
\tilde p_\mu = p_\mu - \frac{1}{4} g g' \rho^2 \theta^3 \delta_{\mu 0}, 
             \nonumber\\
\tilde\rho &=& \rho, ~~~
\tilde\pi_\rho = \pi_\rho +  \frac{1}{2} g^2 \rho A^{0a} \theta^a
                                   -  \frac{1}{2} g g' \rho B^0 \theta^3,
\ea
where 
$U(\theta)=\sum^\infty_{n=0}{{(-1)^n}\over{n!}}({\it ad} \,\theta)^n
=e^{- {\it ad}\,\theta}$.
Therefore, we can obtain the first-class Hamiltonian  
density $\tilde{\cal H}_C$, expressed in terms of the BFT fields as
\ba
\tilde{\cal H}_C &=& {\cal H}_C[\tilde{\cal J}] \nonumber\\ 
\label{18}
 &=& {1 \over 2} (\pi^a_i)^2 
     + {1 \over 2} (p_i)^2
     + {1 \over 2} \left( \pi_\rho +  \frac{1}{2} g^2 \rho A^{0a} \theta^a
                       -  \frac{1}{2} g g' \rho B^0 \theta^3 \right)^2
     + {1 \over 4} (F^a_{ij})^2 
     + {1 \over 4} (G_{ij})^2 
\nonumber\\
&&   + {1\over 8} {\rho}^2 {g'}^2 (B^i)^2
     - {1\over 4} {\rho}^2 g g' B^i  
       \left (U^{3a}(\theta) A^{ia}+V^{3a}(\theta)\partial^i\theta^a \right)
     + {1\over 8} {\rho}^2 g^2  
      \left (U^{ab}(\theta) A^{ib}+V^{ab}(\theta)\partial^i\theta^b\right)^2 
\nonumber\\
&&   +{1\over 2} (\partial_i \rho)^2 - V(\rho)
     + {2 \over {{\rho}^2 g^2}} \left(({\cal D}^i \pi_i)^a\right)^2
     - {2 \over {{\rho}^2 g^2}} (\tilde\Omega^a_2) ^2
\nonumber \\
&&   - B^0 \left\{ {1 \over g} \tilde T_2 - {4 \over {{\rho}^2 g^2}}
               \left ( g' U^{2a}(\theta) ({\cal D}^i \pi_i)^a\tilde\Omega^1_1
                   -g' U^{1a}(\theta) ({\cal D}^i \pi_i)^a\tilde\Omega^2_1
                   \right )
                   \right \}. 
\ea
We moreover observe from these BFT fields 
that the effective first-class constraints (\ref{14}) can be written as
\ba\label{16}
&&\tilde\Omega^a_1=\tilde\pi^a_0,\nonumber \\
&&\tilde\Omega^a_2=(\widetilde{{\cal D}^{i} \pi_i})^a
                   + {1\over 4} {\tilde \rho}^2 g^2 \tilde A^{0a}
                   - {1\over 4} {\tilde \rho}^2 g g' \tilde B^0 \delta^{a3}. 
\ea
Note that comparing with the second-class constraints $\Omega^a_i$
in Eq.(\ref{5}), we see that the constraints  
(\ref{16}) are just the second-class constraints expressed 
in terms of the BFT variables in the extended phase space, showing that
there exists one-to-one mapping between the variables of the reduced and
extended phase spaces. 

As described before, for consistency we should embed the initially 
first-class constraints $T_i$ in Eq.(\ref{6}) as well 
in the extended phase space.
Making use of the BFT fields, these effective first-class constraints can be
easily obtained as
\ba\label{19}
\tilde T_1 
      &=& g \tilde p_0 + g' \tilde \pi^3_0
        = g p_0 + g' \pi^3_0 , \nonumber \\ 
\tilde T_2 
      &=&  g \partial^i \tilde p_i +  g' (\widetilde {{\cal D}^i \pi_i})^3 + 
              {4\over {\tilde \rho^2 g^2}} 
          \left[ g g' (\widetilde {{\cal D}^i \pi_i})^2  \tilde \Omega^1_1  
             - g g' (\widetilde {{\cal D}^i \pi_i})^1  \tilde \Omega^2_1\right]
\nonumber \\
      &=&  g \partial^i p_i +  g' U^{3a} (\theta) ({{\cal D}^i \pi_i})^a  
           +     {4\over {\rho^2 g^2}} 
       \left[g g' U^{2a} (\theta) ({{\cal D}^i \pi_i})^a  \tilde \Omega^1_1  
       - g g' U^{1a} (\theta) ({{\cal D}^i \pi_i})^a \tilde \Omega^2_1\right].
\ea
We do not here need to introduce extra auxiliary fields 
for embedding these first-class constraints. It is only sufficient to use
the BFT fields which are constructed from a pair of the auxiliary fields, 
($\theta,\pi_\theta$).
As results of (\ref{19}), the general properties of 
the first-class constraints are preserved as 
\ba\label{191}
&&\{\tilde T_i(x),\tilde T_j (y)\}=0, \nonumber \\ 
&&\{\tilde \Omega^a_i (x), \tilde \Omega^b_j (y) \}=0, \nonumber \\
&&\{\tilde \Omega^a_i (x), \tilde T_j (y) \}=0.
\ea
We can moreover easily show that the Poisson brackets of the BFT fields
are related with the Dirac brackets of the original fields 
in the unitary gauged SU(2)$\times$U(1) model. 

Therefore, making use of the BFT fields, 
we have shown that the second-class (\ref{5}), the 
first-class constraints (\ref{6}) and the Hamiltonian with their algebraic
relations in the reduced phase space can be replaced with their effective
first-class quantities with the strongly involutive algebraic relations in 
the extended phase space, showing their form invariance in the two
phase spaces. This in general means that when we quantize the system in the 
path integral formulation, due to the strongly involutive relations of the
first-class quantities the partition functional of model can be expressed 
in a simple fashion, {\it i.e.,} there are no terms of $U^a_{bc}$, $V^a_b$
coupled with ghost variables obtained from the usual relations of $\{\Omega^a,
\Omega^b\}=U^{ab}_{c}\Omega^c$, $\{\Omega^a,H\}=V^a_b\Omega^b$. 
Thus, as a result of the BFT construction, we have systematically
obtained the final set of the effective first-class constraints 
$(\tilde \Omega^a_i, \tilde T_j)$ and Hamiltonian $\tilde H_C$
since $U^{ab}_c$, $V^a_b$ appear to be zero in Eq.(\ref{191}).

%%%%%%%%%%%%%%%%%%%%%%%%%%%%%%%%%%%%%%%%%%%%%%%%%%%%%%%%%%%%%%%%%%%%%%%%%%%%%

Now, in order to interpret the results of the BFT construction 
of the effective first-class constraints and Hamiltonian,
let the Lagrangian (\ref{1}) be gauged by making the substitution 
$A^{\mu a}  \rightarrow \hat{A}^{\mu a} 
=  U^{ab}(\theta)A^{\mu b} + V^{ab}(\theta)\partial^\mu \theta^b$.
We then obtain
\be\label{20}
\hat{\cal L}={\cal L}_u+{\cal L}_{WZ},\ee
where
\ba\label{21}
{\cal L}_{WZ}= -{1 \over 4} \rho^2
 g g' B^{\mu} \left( (U^{3a} -\delta^{3a}) A^a_\mu 
                     +V^{3a}  \partial_\mu \theta^a \right)
  + {1 \over 8} \rho^2 g^2 \left (2 V^{ab} A^{\mu a}  
                  \partial_{\mu} \theta^b   
  + V^{ab}  V^{ac} \partial_{\mu} \theta^b
          \partial^{\mu} \theta^c \right)
\ea
plays the role of the Wess-Zumino-Witten (WZW) term like the case of 
the gauge-invariant formulation of two-dimensional chiral gauge
theories \cite{BSV,AAR}.
We then have the momenta $\pi^a_{\theta}$ canonically conjugate to 
${\theta}^a$ as
$\pi^a_{\theta}=- {1 \over 4} \rho^2 g g' B^0 V^{3a}
      + {1 \over 4} \rho^2 g^2 ( U^{bc} A^{0c}  
                + V^{bc} \partial^0 \theta ^c )  V^{ba}$.
The other canonical momenta
$\pi^a_\mu$, $p_\mu$, and $\pi_\rho$ are the same as before. 
Hence the primary constraints
are of the form, $\hat{\Omega}_1^a=\pi_0^a\approx 0$
and $\hat{\Omega}_2=p_0\approx 0$.
The canonical Hamiltonian corresponding to (\ref{20}) reads
\ba\label{22}
\hat {\cal H}_C&=& {1\over 2} (\pi^a_i)^2 
       + {1\over 2} (p_i)^2 
       + {1\over 2} (\pi_\rho)^2
       + {2 \over \rho^2 g^2} (\pi^b_{\theta} (V^{-1})^{ba})^2 
       + {1 \over 4} F^a_{ij} F^{ija}
       + {1 \over 4} G_{ij} G^{ij}
\nonumber \\
&&     + {1\over 8} {\rho}^2 {g'}^2 (B^i)^2
       - {1\over 4} {\rho}^2 g g' B^i  
        \left (U^{3a} A^{ia}+V^{3a} \partial^i\theta^a \right)
       + {1\over 8} {\rho}^2 g^2  
            \left (U^{ab} A^{ib}+V^{ab}\partial^i\theta^b\right)^2 
\nonumber \\
&&    +{1\over 2} (\partial_i \rho)^2 - V(\rho)
      - A^{0a} \left ( ({\cal D}^i \pi_i)^a + 
                \pi^b_{\theta} (V^{-1})^{bc} U^{ca} \right ) 
       - B^0 \left ( \partial^i p_i - 
             {g' \over g} \pi^a_{\theta} (V^{-1})^{a3} \right ),
\ea
where we used the following properties of the Lie algebra-valued functions 
of $U(\theta)$ and $V(\theta)$ \cite{KPR}:
\ba\label{23}
&& U^{ac}(\theta)U^{bc}(\theta)=U^{ca}(\theta)U^{cb}(\theta)
                  =\delta^{ab},~~
          U^{ab}(\theta) \,=\, U^{ba}(-\theta), ~~
          U^{ca}(\theta)V^{cb}(\theta)=V^{ab}(-\theta), \nonumber\\ 
&&        V^{ab}(\theta) = V^{ba}(-\theta),~~ 
          (V^{-1})^{ac} (\theta)  V^{cb} (\theta)  = \delta^{ab}, 
          U^{ac}(\theta)V^{bc}(\theta)\,=\,V^{ab}(\theta),~~
\ea
Persistency in time of the primary constraints
with $\hat H_C$ implies secondary constraints associated with 
the Lagrange multipliers $A^{0a}$ and $B^0$. 
We again have the set of eight first-class constraints as
\ba\label{24}
\hat{\Omega}_1^a &=&\pi_0^a  \approx 0, \nonumber \\
\hat{\Omega}_2 &=& p_0  \approx 0, \nonumber \\
\hat{\Omega}_3^a &=&  ({\cal D}^i \pi_i)^a +
        \pi^b_{\theta} (V^{-1})^{bc} U^{ca} \approx 0, 
        \nonumber \\
\hat{\Omega}_4 &=& \partial^i p_i - 
             {g' \over g} \pi^a_{\theta} (V^{-1})^{a3} \approx 0 .
\ea
We can easily check that all the algebra (\ref{24}) between 
these constraints strongly vanishes except
$\{ \hat{\Omega}_3^a ,\hat{\Omega}_3^b\}=
  g\epsilon ^ {abc} \hat{\Omega}_3^c ~\approx ~ 0$.

It now remains to establish the relation with the BFT results.
We first rewrite the momentum $\pi^{a}_{\theta}$ 
in terms of $\hat {A}^{0a}$ as
$\pi^{a}_{\theta} = - {1 \over 4} \rho^2 g g' B^0 V^{3a} (\theta)
      + {1 \over 4} \rho^2 g^2 \hat{A}^{0b} V^{ba} (\theta)$. 
Making use of Eqs. (\ref{23}) and  (\ref{24}),
we obtain 
\be\label{25}
\hat A^{0a}={4 \over \rho^2 g^2} \left(U^{ab} \hat \Omega^b_3 
              - U^{ab}({\cal D}^i\pi_i)^b
              + {\rho^2 \over 4} g g' B^0 \delta^{a3} \right).
\ee
Comparing this with $A^{0a}$ rewritten by Eqs.(\ref{15}) and (\ref{16}), 
we conclude $\hat A^{0a} \approx \tilde A^{0a}$ since 
$\hat A^{0a}$ and $\tilde A^{0a}$ are identical up to
additive terms proportional to the constraints. 
This establishes the equivalence of $\hat A^{\mu a}$
and $\tilde A^{\mu a}$.

Furthermore, using the matrix $U(\theta)$ and $V(\theta)$ 
we can write the constraints $\hat{\Omega}_3^a \rightarrow
V^{ab}(\theta) \hat{\Omega}^b_3 = V^{ab}(\theta)
({\cal D}^i\pi_i)^b  +\pi^{a}_\theta$ and 
$\hat{\Omega}_4 \rightarrow
g' U^{3a}(\theta) \hat{\Omega}^a_3 + g \hat{\Omega}_4  
= g' U^{3a}(\theta) ({\cal D}^i\pi_i)^a  +g \partial^i p_i$.
Now, performing the canonical transformation \cite{KPR} as
\ba\label{26}
\pi^a_0 &\rightarrow& \pi^a_0 + {1 \over 4} \rho^2 g^2 \theta^a,\nonumber \\
\pi^a_\theta &\rightarrow& \pi^a_\theta + {1 \over 4} \rho^2 g^2 A^{0a}
               -{1 \over 4} \rho^2 g g' B^0 \delta^{a3},   
      \nonumber \\
p_0 &\rightarrow& p_0 - {1 \over 4} \rho^2 g g'\theta^3,\nonumber \\
\pi_\rho &\rightarrow& \pi_\rho + {1 \over 2} \rho g^2 A^{0a}\theta^a
                    - {1 \over 2} \rho g g' B^0 \theta^3,
\ea
we see that the first-class constraints shown in Eq.(\ref{24})
map into the effective constraints (\ref{14}) and (\ref{19})
in the BFT construction as follows
\ba\label{27}
&&\hat{\Omega}^a_1 = \pi^a_0 + {1 \over 4} \rho^2 g^2 \theta^a ~=~  
\tilde{\Omega}^a_1 ,      \nonumber \\
&&V^{ab}(\theta)\hat{\Omega}^b_3 =
  V^{ab}(\theta)
({\cal D}^i\pi_i)^b  +\pi^{a}_\theta + {1 \over 4} \rho^2 g^2 A^{0a}
               -{1 \over 4} \rho^2 g g' B^0 \delta^{a3} ~=~
\tilde{\Omega}^a_2 ,      \nonumber \\
&&g' \hat{\Omega}^3_1 + g \hat{\Omega}_2 =
g' \pi^3_0 + g p_0 ~=~  
\tilde T_1 ,      \nonumber \\
&&g' U^{3a}(\theta) \hat{\Omega}^a_3 + g \hat{\Omega}_4  
=   g' U^{3a}(\theta) ({\cal D}^i\pi_i)^a  +g \partial^i p_i 
~\approx~ \tilde T_2.
\ea
We have thus found that the effective constraints $\tilde{\Omega}^a_2 $ are 
related with the abelian conversion \cite{HT} 
of the constraints $\hat{\Omega}^a_3$
on which the matrix $V(\theta)$ plays a role of converting the non-abelian 
constraints into the abelian ones. 

We can finally check the relation between $\hat {\cal H}_C$
and $\tilde {\cal H}_C$ as given by (\ref{22}) and (\ref{18}).
Making use of $\hat A^{0a}$ and the canonical transformation
(\ref{26}), the expression
(\ref{22}) for $\hat {\cal H}_C$ may be rewritten in the 
following form in order to compare with $\tilde {\cal H}_C$
\ba\label{28}
\hat{\cal H}_C &=& {1 \over 2} (\pi^a_i)^2 
     + {1 \over 2} (p_i)^2
     + {1 \over 2} \left( \pi_\rho +  \frac{1}{2} g^2 \rho A^{0a} \theta^a
                       -  \frac{1}{2} g g' \rho B^0 \theta^3 \right)^2
     + {1 \over 4} (F^a_{ij})^2 
     + {1 \over 4} (G_{ij})^2 
\nonumber \\
&& + {1\over 8} {\rho}^2 {g'}^2 (B^i)^2
     - {1\over 4} {\rho}^2 g g' B^i  
             \left (U^{3a} A^{ia}+V^{3a}\partial^i\theta^a \right)
     + {1\over 8} {\rho}^2 g^2  
            \left (U^{ab} A^{ib}+V^{ab}\partial^i\theta^b\right)^2 
      +{1\over 2} (\partial_i \rho)^2
\nonumber \\
&&      - V(\rho)
        + {2 \over {{\rho}^2 g^2}} \left(({\cal D}^i \pi_i)^a\right)^2
          + {2 \over {{\rho}^2 g^2}} (\hat\Omega^a_3) ^2
          - (A^{0a}+ {4 \over \rho^2 g^2 }({\cal D}^i \pi_i)^a )\hat\Omega^a_3
          - B^0 \hat\Omega_4  . 
\ea
Then, we immediately obtain the equivalence relation 
$\hat {\cal H}_C \approx \tilde {\cal H}_C[\tilde{\cal J}]$ 
since $\hat {\cal H}_C$ is
identical with $\tilde {\cal H}_C[\tilde{\cal J}]$ up to additive terms
proportional to the constraints. 
We thus have arrived at a simple
interpretation of the results obtained in the generalized BFT formalism.
 
%%%%%%%%%%%%%%%%%%%%%%%%%%%%%%%%%%%%%%%%%%%%%%%%%%%%%%%%%%%%%%%%%%%%%%%%%%%

On the other hand, to obtain the corresponding Lagrangian
from the first-class Hamiltonian,
we should perform momenta integrations in the partition functional with 
the delta functionals of the effective 
first-class constraints and proper gauge fixing
functions in the measure \cite{KP,KK}.
However, these procedures could not provide a reasonable result in the 
non-abelian cases \cite{BB,KR,KPR}
because there exist the infinite series terms in some variables such as 
$\tilde A^{0a}$ and $\tilde \Omega^a_2$.
In this paper, we thus suggest a novel path at the classical level
that one can directly get the first-class Lagrangian  
from the original second-class one.
This approach consists in gauging the Lagrangian (\ref{1}), {\it i.e.},
by making use of the BFT fields, $A^{\mu a} \rightarrow \tilde A^{\mu a}$,
$B^\mu \rightarrow \tilde B^\mu$ and $\rho \rightarrow \tilde\rho$.
Since the fields $\tilde B^\mu$ and $\tilde \rho$ in Eq.(\ref{15})
do not have any auxiliary fields, the substitution is trivially performed. 
And the spatial components $\tilde A^{ia}$ of the vector potential
contain only the fields of the configuration space,
and take the usual form of the gauge transformation, {\it i.e.},
$\tilde A^{ia} \rightarrow U^{ab}(\theta) A^{ib}+V^{ab}(\theta)
\partial^i\theta^b$.
However, since $\tilde A^{0a}$ contain the term of 
$\pi_\theta^a$ as in Eq.(\ref{15}), we should first replace this 
with some ordinary fields before carrying out the above substitution.  
This replacement is possible because we have identical relations between
the BFT fields.

From the useful novel property \cite{KK} of 
$\tilde{\cal K} ({\cal J};\theta^a,\pi_\theta^a)
={\cal K} (\tilde{\cal J})$,
where ${\cal K}$, $\tilde{\cal K}$ 
are any second-class and its corresponding effective
first-class function, respectively,
we in particular observe the following relation for $\tilde\pi^a_i$ fields:
\ba\label{29}
\tilde \pi_i^a &=& \partial_i\tilde A_0^a-\partial_0\tilde A_i^a 
                 + g \epsilon^{abc}\tilde A_i^b \tilde A_0^c
\nonumber\\
        &=& \partial_i\left(A_0^a+\frac{4}{g^2 \rho^2}\pi_\theta^a-
            \frac{4}{g^2 \rho^2}\left(U^{ab}-V^{ab}\right)
            ({\cal D}^i\pi_i)^b \right)
            -\partial_0(U^{ab} A_i^{b}+V^{ab}\partial_i\theta^b)
\nonumber\\
        &&   + g \epsilon^{abc}
            (U^{bd} A_i^{d}+V^{bd}\partial_i\theta^d)
            \left(A_0^c+\frac{4}{g^2 \rho^2}\pi_\theta^c-
            \frac{4}{g^2 \rho^2}\left(U^{ce}-V^{ce}\right)
            ({\cal D}^i\pi_i)^e \right). 
\ea
Comparing these with the BFT fields of $\tilde\pi^a_i$ in Eq.(\ref{15}),
{\it i.e.},
$\tilde \pi_i^a =  U^{ab}(\theta)\pi_i^b = 
U^{ab}(\theta) (\partial_i A_0^b-\partial_0 A_i^b 
                 + g \epsilon^{bcd} A_i^c A_0^d)$,
we see that the following
relations should be kept for the consistency 
\be\label{30}                                    
\pi_\theta^a= {1 \over 4} g^2 \rho^2 \left(
    (U^{ab}(\theta)-V^{ab}(\theta)) ({\cal D}^i\pi_i)^b - A^{0a}
     + U^{ab}(\theta)A^{0b} + V^{ab}(\theta)\partial^0 \theta^b    
\right),
\ee
which make it possible to directly rewrite $\tilde A^{0a}$ as 
\be\label{31}
\tilde A^{0a}= U^{ab}(\theta) A^{0b}+V^{ab}(\theta)\partial^0 \theta^b.
\ee
This is the form of the gauge transformation of $A^{0a}$ fields.
As a result, gauging the original Lagrangian (\ref{1}) as
\ba\label{32}
\tilde A^{\mu a} \rightarrow U^{ab}(\theta) A^{\mu b}
+V^{ab}(\theta)\partial^\mu \theta^b,~~
\tilde B^\mu \rightarrow B^\mu, ~~
\tilde \rho \rightarrow \rho,  \ea
we have directly arrived at the first-class Lagrangian as
\be\label{33}
{\cal L}(\tilde A^{\mu a},\tilde B^\mu , \tilde \rho)
=\tilde {\cal L} (A^{\mu a},B^\mu,\theta^a,\rho)
=\hat{\cal L}_{\rm GI}.\ee
Therefore, using the novel relations between the BFT fields,
we can easily obtain the Lagrangian on the space of gauge 
invariant functionals,
and our approach shows that the previous gauging process 
of the Lagrangian in (\ref{20}) make sense.

Finally, by defining the complex scalar doublet 
$\phi (x) = {1 \over \sqrt{2} } e^{ -i g \theta^a (x) {\tau^a \over 2}} 
   {0 \choose \rho (x) +v} = W(\theta) {0 \choose \rho (x) +v}$
with the auxiliary fields $\theta^a$ playing the role of the Goldstone bosons,
and the Pauli matrices $\tau^a~(-i {\tau^a \over 2} = t^a)$, 
we can easily rewrite the Lagrangian (\ref{33}) as 
\be\label{34}
\hat {\cal L}_{GI} = -{1 \over 4} F_{\mu\nu}^a F^{\mu\nu a}
     - {1 \over 4} G_{\mu\nu}G^{\mu\nu} 
     + ({\cal D}_\mu \phi)^\dagger({\cal D}^\mu \phi)
     + V(\phi^\dagger \phi),
\ee
where ${\cal D}_\mu = \partial_\mu - i{g' \over 2}B_\mu 
- i{g \over 2}\tau^a A_\mu^a$.
We therefore have arrived at the symmetric SU(2)$\times$U(1) 
Higgs model from the starting symmetry broken 
Lagrangian (\ref{1}) through the BFT construction.
This proves that the BFT formalism recovers from 
the underlying gauge symmetry of the system.

%%%%%%%%%%%%%%%%%%%%%%%%%%%%%%%%%%%%%%%%%%%%%%%%%%%%%%%%%%%%%%%%%%%%%%%%%%%%

The main objective of this paper was to provide a non-trivial
Hamiltonian embedding of a second-class theory
into a first-class one, following the generalized BFT procedure.
As a result, we have constructed the effective first-class constraints
corresponding to the initially first-class as well as second-class constraints,
and Hamiltonian in particular obtained from the BFT fields.
This process has a great advantage in this non-abelian case 
as compared with the usual BFT approach. 
We have also observed that the effective first-class constraints 
and the Poisson brackets 
of the BFT fields have the form invariance between 
the original and extended phase spaces,
which shows that the BFT fields could be interpreted as the gauge invariant
extension of the second-class variables to the extended phase space.
On the other hand, we have also established the equivalence between 
the effective first-class quantities and corresponding ones 
obtained by gauging the second-class Lagrangian, and
observed that the auxiliary fields $\theta^a$ introduced in the BFT method 
play the role of the Goldstone bosons. 
As a result of the novel relations between the BFT fields,
we have finally obtained
the Lagrangian on the space of gauge invariant functionals, 
which makes it possible to understand the role of the BFT fields
in the configuration space. 
Finally, we hope that through this generalized BFT formalism
the full unbroken symmetry of the universe will be successfully
obtained from the real effective theory describing present symmetry
broken phase.

{\bf Acknowledgment}

Two of the authors (Y.--W. Kim and Y.--J. Park) would
like to thank Prof. K. D. Rothe for his valuable comments and 
warm hospitality at the Institute f\"ur Theoretische Physik. 
The present study was partly supported by  
the Basic Science Research Institute Program, 
the Korea Research Foundation, Project No. 1998-015-D00074.

\end{document}